\documentclass[onecolumn, 12pt]{article}
\usepackage{graphicx}
\usepackage{epsfig}
\usepackage{cite}
\usepackage{amsmath}
\usepackage{biophysj}

\title{Quality Control System Response to Stochastic Growth of Amyloid Fibrils}

\date{\today}

\author{        Simone Pigolotti \\
                Dept. de Fisica i Eng. Nuclear,\\
                Universitat Politecnica de Catalunya\\
                Edif. GAIA, Rambla Sant Nebridi s/n,\\
                08222 Terrassa, Barcelona, Spain\\
                \and Ludvig Lizana\\ 
		Niels Bohr Institute,\\
		Blegdamsvej 17, DK-2100 Copenhagen, Denmark\\
		\and Daniel Otzen \\
		University of Aarhus,\\
                iNANO, Department of Molecular Biology and Genetics\\
                Center for Insoluble Protein Structures (inSPIN) \\
                Gustav Wieds Vej 10C Aarhus C, Denmark\\
		\and Kim  Sneppen\\ 
		Niels Bohr Institute,\\
		Blegdamsvej 17, DK-2100 Copenhagen, Denmark\\
		}

\begin{document}
\maketitle


\begin{abstract}
  We introduce a stochastic model describing aggregation of misfolded
  proteins and degradation by the protein quality control system in a
  single cell. In analogy with existing literature, aggregates can
  grow, nucleate and fragment stochastically. We assume that the
  quality control system acts as an enzyme that can degrade aggregates
  at different stages of the growth process, with an efficiency that
  decreases with the size of the aggregate. We show how this
  stochastic dynamics, depending on the parameter choice, leads to two
  qualitatively different behaviors: a homeostatic state, where the
  quality control system is stable and aggregates of large sizes are
  not formed, and an oscillatory state, where the quality control
  system periodically breaks down, allowing for the formation of large
  aggregates. We discuss how these periodic breakdowns may constitute
  a mechanism for the sporadic development of neurodegenerative
  diseases.  \emph{Key words: protein aggregation | Neurodegenerative
    diseases | stochastic dynamics | spiky oscillations | proteasome | lysosome}
\end{abstract}


\section{Introduction}

Protein aggregation and formation of amyloid fibrils have been a
subject of intense experimental and theoretical study in recent
years. The main motivation has been that protein aggregation,
including the formation of cytotoxic pre-amyloid species and larger
inclusion bodies, seems to be the common theme underlying most known
neurodegenerative diseases \cite{ross,chiti}.  Besides structural
studies, several attempts have been made to build up coarse-grained
models, able to characterise the aggregation process from a kinetic
point of view without any influence of a cellular environment.  One of
the key features which are needed in the description is the existence
of a nucleation mechanism \cite{ferronejames}. During the late stages
of the aggregation, the possibility of aggregates to break into
smaller fragments also becomes important
\cite{poschel,collins}. Several other models have later been 
introduced, including more detailed mechanisms
\cite{kunes,knowles_science,cabriolu}.

While amyloid growth and aggregation {\it in vitro} is quite well
characterized nowadays, much less is known about the corresponding
phenomenon {\it in vivo}. This is the main focus in this paper. 
The {\it in vivo} case is of crucial relevance since a common
theme of many neurological disorders is undesired intercellular
protein aggregation, and insights would be thus helpful to the
understanding of disease development as well as treatments.

Apart from the experimental limitations, the {\it in vivo} case
presents also a number of theoretical challenges. Accepting as a
working hypothesis that the development of neurological disorders is
directly associated with the aggregation process, it is unclear why
such diseases develop in a time frame of several years, while {\it in
  vitro} aggregates can typically grow on timescales of hours or
days. Another ubiquitous feature of neurological disorder is their
sporadic development: degeneration is not gradual but occurs in
almost-discrete steps. While a similar dynamical behaviour has been
observed {\it in vitro} \cite{kellermayer,fonslet}, also in this case
there is a large gap between the timescales observed in experiments
and those characterizing the disease.

We examine the dynamics of {\it in vivo} aggregation where the cell's
quality control system tries to prevent protein aggregation by acting
at different stages of the growth process. For example, in the case of
Parkinson disease, it has been observed how $\alpha$-synuclein can be
degraded both by the ubiquitin-proteasome system (UPS) and by
autophagy \cite{webb}. The possibility of such mechanisms to degrade
aggregates of $\alpha$-synuclein is currently under debate.

In a previous study, we explored the consequences of the interaction
between $\alpha$-synuclein and the UPS system \cite{SLJPO}. We found
that such system displays a transition between a state in which the
UPS system can effectively prevent aggregation, and one in which spiky
oscillations are observed. During such spikes, the UPS  is
impaired, allowing for growth of aggregates. However, the model we
considered lacked a description of the aggregation kinetics.

To address this issue, we study a model describing the behaviour of
the battle between protein aggregation and the quality control system
in a single cell. We implement the aggregation process in a similar
way as in \cite{knowles_science}: aggregates are formed when a
nucleation threshold is passed, and grow according to the existing
concentration of monomers. Further, large aggregates can break into
smaller fragments.  To this mechanism, we couple the response of the
protein quality control system. We assume that its action can be
modeled as an enzymatic degradation, whose efficiency depends on the
size of the aggregate. We mostly identify this system with the UPS, as
it is better characterized from a biochemical point of view than other
systems such as lysosomes. However, larger fragments are more likely
to be attacked by autophagy, whose possible effect we will also
discuss.

In order to assess the effect of intrinsic noise on the resulting
dynamics, we have implemented all reactions stochastically using the
Gillespie algorithm \cite{gillespie}.  We observe that, even in the
presence of stochasticity, aggregation in this model is not
continuous, but occurs in fairly regular spikes. During these spikes,
aggregates of all sizes can quickly be formed. Such spikes become more
irregular if the degrading agent is present on average in low numbers
in the cell. Before discussing the consequences of our findings, we
outline the details of our model.


\section{Model}

\begin{figure*}
\includegraphics[width= \textwidth]{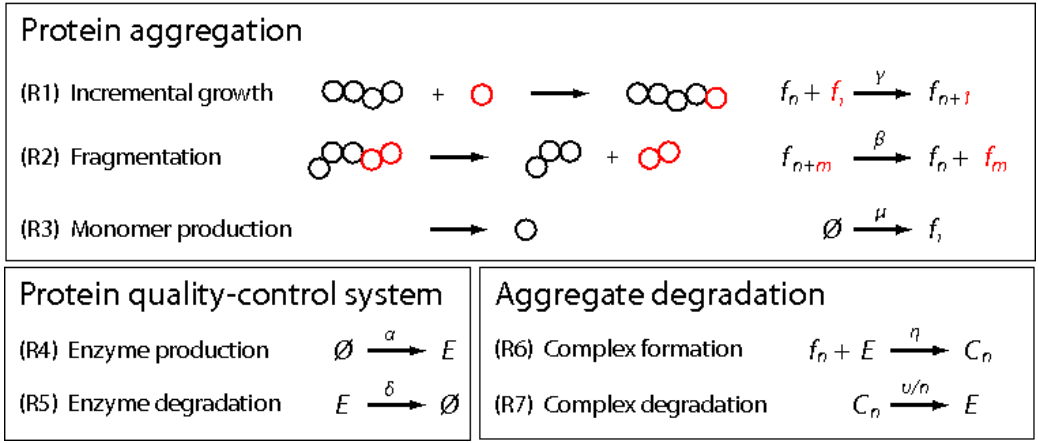}
\caption{Total set of reactions that defines our model of protein
  aggregation (top panel) and subsequent degradation by the cell's
  quality control system (bottom right panel). The quality-control
  system is under cellular regulation (bottom left panel).}
\label{fig:model}
\end{figure*}

Our model describes the stochastic growth of fibrillar aggregates
during constant attack by the cell's protein quality-control
system. We define $f_n$ as the number of free aggregates made up of
$n$ monomers present in the cell, so that $f_1$ is the number of free
monomers, or aggregation-prone proteins. We will assume that the action of
the quality control system can be modelled as an enzymatic
degradation, where we have the case of the UPS response system as an
example in mind. The number of free degradation enzymes available in
the cell is denoted by $E$. Such enzymes can degrade fibrils of size
$n $ through the formation of a long lived complex $C_n$. The
interplay between $f_n, E$ and $C_n$ is summarised in
Fig. (\ref{fig:model}) and discussed below.

In analogy with existing models \cite{poschel,knowles_science} we
describe the dynamics of aggregate formation in terms of two basic
processes (top panel in Fig. \ref{fig:model}): (R1) step-wise growth
by attachment of monomers to one of the fibril ends, and (R2) random
breakage of aggregates into smaller fragments. Reactions (R1) and (R2)
occur at rates $\gamma$ and $\beta$, respectively. In addition to (R1)
there is a possibility to introduce a nucleation barrier so that the
reaction in which two monomers bind (R1') $f_1+f_1 \xrightarrow{\gamma'}
f_2$ is slower than (R1), i.e.  $\gamma'\ll \gamma$. The dynamics of
(R1) and (R2) have been shown to be consistent with results of {\it in
  vitro} aggregation experiments \cite{knowles_science}. In addition to
(R1) and (R2), we consider a constant production of monomers at rate
$\mu$ (R3).

The cell's protein quality-control system, here represented by a
generic degradation enzyme $E$, attacks fibrillar aggregates of all
sizes in order to keep their concentration at a low level. The enzyme
is produced at rate $\alpha$ (R4) and each molecule is degraded with
rate $\delta$ (R5) (bottom left panel). The steady-state
concentration of $E$ in absence of fibrils is then $\alpha/\delta$.

The degradation of aggregates by the degradation enzyme is assumed to
take place as a two-step process (bottom right panel). First, it binds
to fibrils of a given size $n$ and forms a complex $C_n$ (R6). Then,
in a time proportional to $n$, the aggregate is entirely degraded and
the enzyme is released and ready to deal with another aggregate (R7).
In general, it is unreasonable to assume that the degradation
capability of the quality control is simply inversely proportional to
$n$. A more realistic assumption would be to let $\eta$ and $\nu$ have
a non-trivial size dependence. The functions $\eta(n)$ and $\nu(n)$
may be also different for different agents, describing the spectrum of
aggregates that each of them can effectively bind and degrade.  For
example, a proteasome could be able to attack misfolded monomers and
eventually early stage aggregates, while lysosomes might effectively
attack larger bodies. However, given the poor biochemical
characterisation of these agents, we consider only one of them at a
time and keep $\eta$ and $\nu$ independent of $n$. However, to avoid
the possibility of the enzyme to attack unreasonably large aggregates,
we introduce an upper limit $n_{\rm max}$ to the aggregate size that
the proteasome can bind to.

In summary, our model is defined by reactions
(R1)-(R7) with ingoing parameters $\gamma$($\gamma'$), $\beta$, $\mu$,
$\alpha$, $\delta$, $\eta$, $\nu$ and $n_{\rm max}$.


\section{Results}

We investigated the dynamics of the model introduced in the previous
section using computer simulations. In order to include properly the
effects caused by intrinsic noise, we implemented all the reactions
stochastichally via the Gillespie algorithm \cite{gillespie}. 
Parameters were selected according to physiological conditions based
on measured aggregation rates of proteasomal dynamics. The one
exception is $\alpha$ which is left as a free parameter in our
model. Since we are interested in the long-time behaviour, all time
units are in days. Reactions rates are therefore expressed in
$[day]^{-1}$, $[day]^{-1}$ per molecule or $[day]^{-1}$ per couple of
molecules, depending on reaction order (number of chemical species on
the reaction's left-hand side).  We will start our analysis by
identifying the enzyme with the ubiquitin-proteasome system, which we
refer to simply as the proteasome, the main reason being that it is
biochemically well characterised (compared to e.g. lysosomes).

The parameters characterizing the aggregating monomers have been
chosen having $\alpha$-synuclein in mind as example. However, the
model can be easily adapted to describe aggregation of other proteins.
We fix the parameters according to this choice and later vary key
parameters in order to probe the different behaviours of the
model. The specific choice of all numerical values of the parameters
is discussed in Appendix \ref{apppar}.

\begin{figure}[htb]
\center
\includegraphics[width=0.9\textwidth]{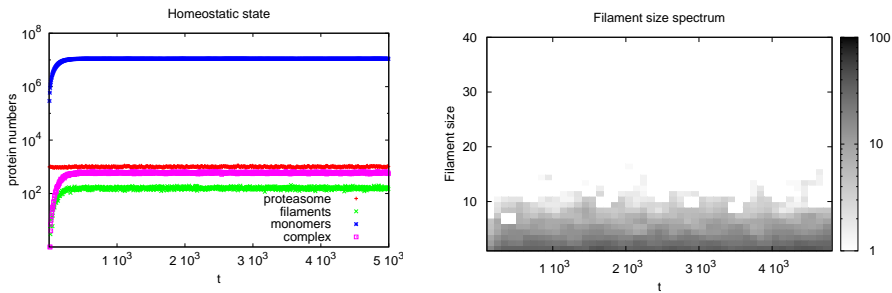}
\caption{Homeostatic state. (left) Number of molecules as a function
  of time (right) size distribution of bound aggregates. The following
  parameters were used: $\gamma=2\cdot 10^{-5}$, $ \gamma'=2\cdot
  10^{-10}$, $\beta=10^{-3}$, $n_{\rm max}=10^{6}$, $\mu=6\cdot 10^4$,
  $\alpha=10^2$, $\delta=0.1$, $\eta=0.02$, $\nu=10^2$. All rates are
  in unit $[day]^{-1}$, $[day]^{-1}$ per molecule or $[day]^{-1}$ per
  couple of molecules, depending on the order of the reaction.}
  \label{nonoscill}
\end{figure}

Figure \ref{nonoscill} shows the dynamics of the model with the
parameter choice from Appendix \ref{apppar} without any
alteration. The result is a stationary homeostatic state, after a
short period of transient behaviour from the initial conditions, where
there is a chemical balance between proteasome concentration and the
formation of aggregates. The proteasome is in the model denoted $E$
whereas filaments and bound filaments represent a sum over all sizes
excluding the free monomers $f_1$, that is $\sum_{n>1}f_n$ and $\sum_n
C_n$. The figure shows that their concentrations fluctuate moderately
around well defined average values which means that the proteasome is
fully functioning and keeps the total aggregate concentration down. In
this homeostatic state there is also a fast decay (well fitted by an
exponential) of the aggregate size distribution as a
function of $n$ (right panel) as well as bound fibrils ($C_n$) (not
shown).

Coupled kinetic equations for the aggregation process (R1)-(R2) with a
finite pool of monomers were solved analytically in
\cite{knowles_science}. The model we consider here is more complicated
due to the non-linear coupling with the proteasome and does therefore
not lend itself easily to analytical treatments. The problem does
however simplify if we let the aggregate size $n$ be a continuous
variable. Indeed, one can show (see Appendix \ref{appmath}) that the
aggregate size-distribution in the homeostatic case decays
exponentially with $n$.  This means that very large fibrillar
aggregates are unlikely to be formed in this regime; as a consequence,
the value of $n_{\rm max}$ is irrelevant, as soon as it is reasonably
large. This contrasts the behaviour of the model in the absence of any
proteasome degradation where fibrils grow at a much higher rate, up to
a size which is essentially unlimited.  An unlimited growth that in
{\it in vitro} experiments becomes limited by the available pool of
aggregating peptide chains \cite{knowles_science}.

\begin{figure}[htb]
\center
\includegraphics[width=0.9\textwidth]{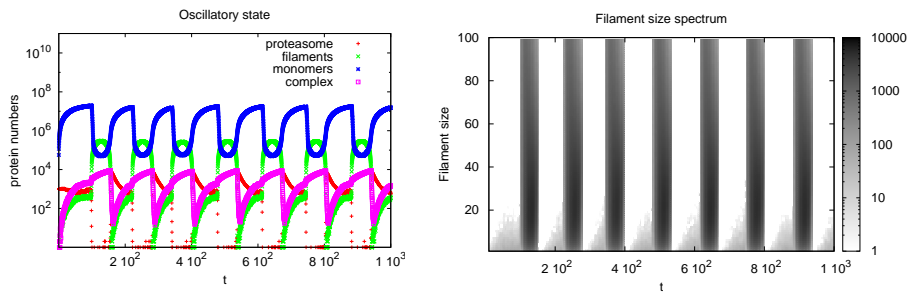}
\caption{Oscillatory state, number of molecules as a function of
  time. Parameters are same as Fig. \ref{nonoscill} but $\mu$ is 5 times
  larger, $\mu=3\cdot 10^5$. (left) number of molecules as a function
  of time, (right) size distribution of bound aggregates.}
  \label{oscill}
\end{figure}

If the monomer production rate ($\mu$) is increased, homeostasis
breaks down and oscillations emerge (Fig. \ref{oscill}, left
panel). In this regime there are dramatic drops in proteasome
concentration which stays low for long periods of time, leading to
a buildup of larger and larger fibrillar aggregates. The quality
control system cannot in this case cope with the amount of aggregates
present in the system which results in a non-exponential decay in the
aggregate size distribution with $n$. The distribution becomes heavy
tailed, and aggregates may reach sizes larger than $n_{\rm max}$,
i.e. large enough to escape the repair system.

The striking outcome of the model is that destabilisation of the
quality control system is episodic. The system alternates between
states in which the quality control system is functioning, and short
periods in which the proteasome is impaired. During such periods,
aggregates of all sizes are predicted, as shown in the right panel of
Fig. \ref{oscill}.  This is a consequence of the fact that the growth
dynamics, in the absence of the proteasome, is faster than the timescale
of the recovery of the quality control system. The largest aggregates
size that the proteasome can attack is controlled by $n_{\rm max}$ and
if aggregates larger than $n_{\rm max}$ are formed, nothing can
prevent such aggregates from reaching very large sizes, even when the
quality control system has recovered.  Finally, despite the fact that
simulations are stochastic, the period of the oscillations is
remarkably regular. This is a consequence of the (average) high copy
number of all proteins involved.

We also remark that, starting from the homeostatic state of
Fig. \ref{nonoscill}, the onset of oscillations can be triggered by
varying different parameters. In particular, reducing the proteasome
production rate $\alpha$, and/or the proteasome degradation efficiency
$\nu$ trigger oscillations. 

\subsection{Lysosomal degradation}

Autophagy is the process by which cytosolic membrane-bound
compartments engulf substrates, such as mis-folded proteins or protein
aggregates, that ultimately fuse with the lysosome for degradation of
their content \cite{tyedmers}. For example, mis-folded
$\alpha$-synuclein protein in Parkinson's disease has been identified
as substrate for this type of autophagy \cite{webb}. Autophagy can be
regarded as a back-up system to complement proteasomal degradation
when it is overwhelmed or incapable of dealing with specific
aggregated substrates. Indeed, experiments strongly suggest that the
removal of mis-folded proteins through autophagy and the ubiquitin
proteasome system are interconnected \cite{pandey}, as impairment of
the ubiquitin proteasome system induces compensatory autophagy.

\begin{figure}[htb]
\center
\includegraphics[width=0.9\textwidth]{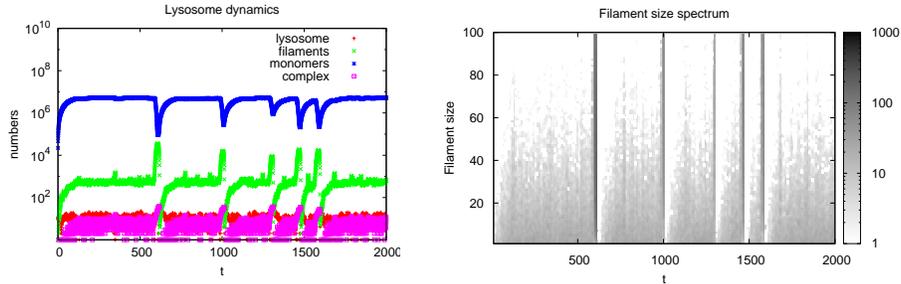}
\caption{Oscillations in the presence of lysosome. Parameters are:
  $\gamma=2\cdot 10^{-5}$,$\gamma'=2\cdot 10^{-10}$, $\beta=10^{-3}$,
  $n_{\rm max}=10^6$, $\mu=6\cdot 10^4$, $\alpha=1$,$\delta=0.1$,
  $\eta=0.02$, $\nu=10^4$. The only parameters that differ from the
  case of Fig. \ref{nonoscill} are $\alpha$ and $\nu$.  As usual,
  (left) number of molecules as a function of time, (right) size
  distribution of filaments. Notice how oscillations are much more
  stochastic as a consequence of the presence of small numbers of
  lysosomes.}
  \label{fig_lyso}
\end{figure}

This type of protein degradation is not so well characterised
biochemically compared to for instance proteasomal degradation. But in
order to test the effect of autophagic-like degradation we tuned our
model parameters to represent the lysosome as an efficient enzyme
present only in small copy numbers. We assumed a production rate
$\alpha=1$ lysosome per day, and a degradation rate of $\delta=0.1$
lysosomes per day which corresponds to an average of $10$ lysosomes
per cell at steady state. We assumed furthermore that the rate at
which the enzyme degrades aggregates (per aggregate size) is
$\nu=10^4$, which is a hundred times more efficient than the
proteasome. Other mechanisms may be introduced to describe lysosomal
dynamics in a more realistic ways. For example, one could introduce
the possibility to deal with multiple aggregates at the same time, or
a processing time having a different dependence on the $n$. For
simplicity, we do not explore such possibilities here.

Figure \ref{fig_lyso} shows the results of a simulation for our new
parameter choice. Even in this case we observe oscillations: the
reason is that the higher efficiency is compensated by the lower
numbers. The main difference between this case and that in
fig. \ref{oscill} is that the period of the oscillations is much more
stochastic. This is a consequence of the enzyme being present in small
numbers and the breakdown of the protein quality control system is
much less regular in this case.


\section{Discussion}

In this paper we have studied the interplay between sequential growth
of amyloid fibrils and its degradation by the quality control system
in a single cell. Our model exhibit two types of dynamical
behaviours. For one set of parameters the number of molecules is
stationary and fluctuate around a well defined constant mean, while
for the complementary set the system oscillates. These features were
also captured in a simplified deterministic model \cite{SLJPO} where
the incremental fibrillar growth was replaced by a two-step process
(small and large aggregates). However, the present study demonstrates
how the dynamical behaviour observed in \cite{SLJPO} is present also
in a more mechanistic model, where the aggregation process is taken
explicitly into account. Further, we demonstrate how such transition
can occur for parameters values, to the best of our knowledge, under
physiological conditions.

The main result of the model is that, in the regime in which the
quality control system can not cope with the amount of fibrils,
destabilisation of the homeostatic state gives rise to an oscillatory
regime. Such oscillations are characterised by long time lapses in
which the quality control system is functioning and prevents
aggregation, separated by shorter lapses of time in which the quality
control system is impaired and fibrils of large sizes can grow. The
period of such oscillations depends on the parameter choice but is
typically on the scale of months, thus predicting a slow and stepwise
formation of large aggregates. Such phenomenology is reminiscent of the
remarkably slow and sporadic development of neurodegenerative
diseases.

Finally, we have shown that when the agent degrading the aggregates is
present in large numbers, the resulting oscillations tend to be very
regular and almost deterministic, while when the numbers are smaller
oscillations are more stochastic.

There are clearly a number of generalisations one can consider to
account for more accurate experimental evidences, such as the
possibility of different interconnected repair systems, that attack
aggregates at different stages of the growth process and with
different efficiencies depending on the aggregate size. Our model
should be considered as a minimal mechanistic model displaying such
non-trivial phenomenology.


\section{Acknowledgemens}
This work was funded by the Danish National Research Foundation
through the Center for Models of Life (CMOL). L.L acknowledges the
financial support by the Knut and Alice Wallenberg
foundation. D.E.O. is funded by the Danish Research Foundation
(inSPIN) and the Lundbeck Foundation (BioNET 2).


\section{Appendix: choice of parameters}\label{apppar}

In this appendix, we discuss our initial choice of parameters. We
start the discussion with the parameters characterizing the
aggregation/fragmentation part of the model. Studies on insulin
\cite{knowles_pnas} reported fibril growth rates around $0.3$ monomers
per fibril per second in the presence of a monomer concentration of
$0.17\ mM$. Similar studies on other molecules \cite{kusumoto} report
similar values.  The average diameter of a dopaminergic neuron is from
$10$ to $20 \mu m$ (see e.g. \cite{hout}). This will give a cell
volume around $v=10^{-11}l$. One single monomer in the cell will then
correspond to a concentration of $1/(N_Av)\approx 2\cdot 10^{-13} M$
where $N_A$ is the Avogadro number. The growth rate per fibril per
monomer will then be $0.25\cdot 10^{-9} s^{-1}$, that is approximately
$2\cdot 10^{-5}$ monomers per fibril per day. In
Ref. \cite{knowles_pnas} it is estimated from reaction rate arguments
that one every $10^5$ encounter events between filaments and monomer
result in attachment.

Fitting models on aggregation to experiments suggest that the
nucleation barrier can lead to an activation rate being $8$-$9$ orders
of magnitudes smaller than the growth rate after nucleation
\cite{knowles_science}. This rate could however be different in vivo,
since cell condition can significantly alter the nucleation barrier
and favour/disfavour the formation of nucleation seeds. We choose a
relatively high nucleation rate (compared with in vitro experiments),
$\gamma'= 2\cdot 10^{-10}$.

We fixed a filament breakage rate $\beta=10^{-3}$
\cite{knowles_science}. We observed that varying this rate has little
effect on the outcome of the model, as most filaments are actively
degraded by the quality control system. We also fixed $n_{\rm
  max}=10^6$; also this last parameter do not change the qualitative
dynamics of the system when it is large enough, as the distribution of
fibrils decays exponentially with their size in the stable
regime. However, lowering the value will increase the chance that an
aggregate may grow large enough to escape degradation to the quality
control system. The production rate of monomers $\alpha$ is left as a
free parameters. Since its value is typically large, to speed up the
simulations we implemented an alternative reaction in which $100$
monomers at a time are produced with a rate $\alpha/100$. We checked
in shorter simulations that such approximation does not alter
significatively the results, thanks to the typical high numbers of
monomers.

We move now to the parameters related to the proteasome. Proteasome
lifetime is estimated to be 8-15 days in the cell
\cite{tanaka,cuervo}, leading to a degradation rate $\delta=0.1$. The
proteasome production rate $\alpha$ is left as a free parameter. On
general grounds, we may expect the proteasome to bind better to the
filaments than to a monomer, leading to a complex formation $\eta$
being larger than $\gamma$. We pick $\eta=0.02$, corresponding to one
successful binding event every $100$ encounters. A lower value of
$\eta$ would make the proteasome less effective and would simply
result in a higher threshold value for $\alpha$ to observe
oscillations, without affecting the qualitative features of the
model. Finally, the degradation rate of individual molecules by the
proteasome is estimated to be on the order of minutes, so we pick
$\nu=10^3$.

\section{Appendix: continuous limit}\label{appmath}

If we treat the number of number of monomers $n$ forming an aggregate
as a continuous variable, our model can be treated
analytically. Adopting the notation from Fig. \ref{fig:model} the
governing equations for the molecular concentrations are
\begin{eqnarray}
\frac{\partial f_n(t)}{\partial t} &=& - 
\gamma \frac{\partial f_n(t)}{\partial n} 
-\eta f_n(t) E(t), 
\\
\frac{\partial C_n(t)}{\partial t} &=& \eta f_n(t) E(t) - \frac\nu n C_n(t),
\\
\frac{\partial E(t)}{\partial t} &=& 1-E(t) + 
\int_1^{n_{\rm max}} dn \left(  \eta f_n(t) E(t) - \frac \nu n C_n(t) \right),
\end{eqnarray}
Here we neglected filament breakage ($\beta=0$) and $n\geq1$. We have
also rescaled enzymatic growth and degradation rates as well as the
filament length $n$ to 1. The constant production of monomers enters
as the boundary condition
\begin{equation}
        -\left(\frac{\partial f_n(t)}{\partial n}\right)_{n=1} = \mu.
\end{equation}

In the homeostatic "healthy" state it is straightforward to
show that the time independent stationary (ST) concentrations are
given by
\begin{eqnarray}
        f_n^{\rm ST}(n) = \frac {\mu \gamma}\eta e^{-\eta(n-1)/\gamma }, \ \ \ \
        C_n^{\rm ST}(n) = \frac {n\gamma\mu}\mu {\nu} e^{-\eta(n-1)/\gamma }, \ \ \ \ 
        P^{\rm ST} = 1,
\end{eqnarray}
which shows that the size distribution of filaments $f_n^{\rm ST}$
decays exponentially. Moreover, the complex concentration $C_n^{\rm
  ST}$ grows with $n$ until it reaches a maximum at $\bar n=\eta/\gamma$
after which it also decays exponentially.

\end{document}